\documentclass[%
hidelinks,
reprint,
superscriptaddress,
showpacs,
amsmath,amssymb,
prc,
altaffilletter
]{revtex4-2}

\usepackage{graphicx}
\usepackage{dcolumn}
\usepackage{bm}
\usepackage{orcidlink}
\usepackage{upgreek}
\usepackage{mathrsfs}

\def\MJDEMit{\itshape{\scshape{Majorana Demonstrator}}}
\def\MJit{\itshape{\scshape{Majorana}}}
\def\DEMit{\itshape{\scshape{Demonstrator}}}

\newcommand{\MaGe}{\textsc{MaGe}}
\def\qval{$Q_{\beta\beta}$}  
\def\novbb{$0\nu\beta\beta$}
\def\twovbb{$2\nu\beta\beta$}
\def\Ge{$^{76}$Ge}
\def\Th{$^{232}$Th}
\def\Ur{$^{238}$U}

\begin{document}

%\preprint{AAPM/123-QED}

\title{An assay-based background projection for the {\sc{Majorana Demonstrator}} using Monte Carlo Uncertainty Propagation}

\newcommand{\ITEP}{National Research Center ``Kurchatov Institute'', Kurchatov Complex of Theoretical and Experimental Physics, Moscow, 117218 Russia}
\newcommand{\JINR}{Joint Institute for Nuclear Research, Dubna, 141980 Russia} 
\newcommand{\lbnl}{Nuclear Science Division, Lawrence Berkeley National Laboratory, Berkeley, CA 94720, USA}
\newcommand{\lbnle}{Engineering Division, Lawrence Berkeley National Laboratory, Berkeley, CA 94720, USA}
\newcommand{\lanl}{Los Alamos National Laboratory, Los Alamos, NM 87545, USA}
\newcommand{\queens}{Department of Physics, Engineering Physics and Astronomy, Queen's University, Kingston, ON K7L 3N6, Canada}
\newcommand{\uw}{Center for Experimental Nuclear Physics and Astrophysics, and Department of Physics, University of Washington, Seattle, WA 98195, USA}
\newcommand{\unc}{Department of Physics and Astronomy, University of North Carolina, Chapel Hill, NC 27514, USA}
\newcommand{\duke}{Department of Physics, Duke University, Durham, NC 27708, USA}
\newcommand{\ncsu}{Department of Physics, North Carolina State University, Raleigh, NC 27695, USA}	
\newcommand{\ornl}{Oak Ridge National Laboratory, Oak Ridge, TN 37830, USA}
\newcommand{\ou}{Research Center for Nuclear Physics, Osaka University, Ibaraki, Osaka 567-0047, Japan}
\newcommand{\pnnl}{Pacific Northwest National Laboratory, Richland, WA 99354, USA}
\newcommand{\ttu}{Tennessee Tech University, Cookeville, TN 38505, USA}
\newcommand{\sdsmt}{South Dakota Mines, Rapid City, SD 57701, USA}
\newcommand{\usc}{Department of Physics and Astronomy, University of South Carolina, Columbia, SC 29208, USA}
\newcommand{\usd}{Department of Physics, University of South Dakota, Vermillion, SD 57069, USA}  
\newcommand{\ut}{Department of Physics and Astronomy, University of Tennessee, Knoxville, TN 37916, USA}
\newcommand{\tunl}{Triangle Universities Nuclear Laboratory, Durham, NC 27708, USA}
\newcommand{\mpi}{Max-Planck-Institut f\"{u}r Physik, M\"{u}nchen, 80805, Germany}
\newcommand{\tum}{Physik Department and Excellence Cluster Universe, Technische Universit\"{a}t, M\"{u}nchen, 85748 Germany}
\newcommand{\williams}{Physics Department, Williams College, Williamstown, MA 01267, USA}
\newcommand{\ciemat}{Centro de Investigaciones Energ\'{e}ticas, Medioambientales y Tecnol\'{o}gicas, CIEMAT 28040, Madrid, Spain}
\newcommand{\iu}{IU Center for Exploration of Energy and Matter, and Department of Physics, Indiana University, Bloomington, IN 47405, USA}

\newcommand{\presentaddressllnl}{Present address: Lawrence Livermore National Laboratory, Livermore, CA 94550, USA}

\author{I.J.~Arnquist}\affiliation{\pnnl} 
\author{F.T.~Avignone~III}\affiliation{\usc}\affiliation{\ornl}
\author{A.S.~Barabash\,\orcidlink{0000-0002-5130-0922}}\affiliation{\ITEP}
\author{C.J.~Barton}\affiliation{\usd}	
\author{K.H.~Bhimani}\affiliation{\unc}\affiliation{\tunl} 
\author{E.~Blalock}\affiliation{\ncsu}\affiliation{\tunl} 
\author{B.~Bos}\affiliation{\unc}\affiliation{\tunl} 
\author{M.~Busch}\affiliation{\duke}\affiliation{\tunl} 
\author{T.S.~Caldwell}\affiliation{\unc}\affiliation{\tunl}	
\author{Y.-D.~Chan}\affiliation{\lbnl}
\author{C.D.~Christofferson}\affiliation{\sdsmt} 
\author{P.-H.~Chu\,\orcidlink{0000-0003-1372-2910}}\affiliation{\lanl} 
\author{M.L.~Clark}\affiliation{\unc}\affiliation{\tunl} 
\author{C.~Cuesta\,\orcidlink{0000-0003-1190-7233}}\affiliation{\ciemat}	
\author{J.A.~Detwiler\,\orcidlink{0000-0002-9050-4610}}\affiliation{\uw}	
\author{Yu.~Efremenko}\affiliation{\ut}\affiliation{\ornl}
\author{H.~Ejiri}\affiliation{\ou}
\author{S.R.~Elliott\,\orcidlink{0000-0001-9361-9870}}\affiliation{\lanl}
\author{N.~Fuad\,\orcidlink{0000-0002-5445-2534}}\affiliation{\iu}
\author{G.K.~Giovanetti}\affiliation{\williams}  
\author{M.P.~Green\,\orcidlink{0000-0002-1958-8030}}\affiliation{\ncsu}\affiliation{\tunl}\affiliation{\ornl}   
\author{J.~Gruszko\,\orcidlink{0000-0002-3777-2237}}\affiliation{\unc}\affiliation{\tunl} 
\author{I.S.~Guinn\,\orcidlink{0000-0002-2424-3272}}\affiliation{\ornl} 
\author{V.E.~Guiseppe\,\orcidlink{0000-0002-0078-7101}}\affiliation{\ornl}	
\author{C.R.~Haufe}\affiliation{\unc}\affiliation{\tunl}	
\author{R.~Henning}\affiliation{\unc}\affiliation{\tunl}
\author{D.~Hervas~Aguilar\,\orcidlink{0000-0002-9686-0659}}\altaffiliation{Present address: Technical University of Munich, 85748 Garching, Germany} \affiliation{\unc}\affiliation{\tunl}
\author{E.W.~Hoppe}\affiliation{\pnnl}
\author{A.~Hostiuc}\affiliation{\uw} 
\author{M.F.~Kidd}\affiliation{\ttu}	
\author{I.~Kim}\altaffiliation{\presentaddressllnl} \affiliation{\lanl}
\author{R.T.~Kouzes}\affiliation{\pnnl}
\author{T.E.~Lannen~V}\affiliation{\usc} 
\author{A.~Li\,\orcidlink{0000-0002-4844-9339}}\affiliation{\unc}\affiliation{\tunl} 
\author{J.M. L\'opez-Casta\~no}\affiliation{\ornl} 
\author{R.D.~Martin,\orcidlink{0000-0001-8648-1658}}\affiliation{\queens}	
\author{R.~Massarczyk}\affiliation{\lanl}		
\author{S.J.~Meijer\,\orcidlink{0000-0002-1366-0361}}\affiliation{\lanl}	
\author{T.K.~Oli\,\orcidlink{0000-0001-8857-3716}}\altaffiliation{Present address: Argonne National Laboratory, Lemont, IL 60439, USA}\affiliation{\usd}
\author{L.S.~Paudel\,\orcidlink{0000-0003-3100-4074}}\affiliation{\usd} 
\author{W.~Pettus\,\orcidlink{0000-0003-4947-7400}}\affiliation{\iu}	
\author{A.W.P.~Poon\,\orcidlink{0000-0003-2684-6402}}\affiliation{\lbnl}
\author{D.C.~Radford}\affiliation{\ornl}
\author{A.L.~Reine\,\orcidlink{0000-0002-5900-8299}}\affiliation{\unc}\affiliation{\tunl}	
\author{K.~Rielage\,\orcidlink{0000-0002-7392-7152}}\affiliation{\lanl}
\author{N.W.~Ruof\,\orcidlink{0000-0002-0477-7488}}\altaffiliation{\presentaddressllnl} \affiliation{\uw}
\author{D.C.~Schaper}\affiliation{\lanl} 
\author{S.J.~Schleich\,\orcidlink{0000-0003-1878-9102}}\affiliation{\iu}
\author{D.~Tedeschi}\affiliation{\usc}		
\author{R.L.~Varner\,\orcidlink{0000-0002-0477-7488}}\affiliation{\ornl}  
\author{S.~Vasilyev}\affiliation{\JINR}	
\author{S.L.~Watkins\,\orcidlink{0000-0003-0649-1923}}\affiliation{\lanl}
\author{J.F.~Wilkerson\,\orcidlink{0000-0002-0342-0217}}\affiliation{\unc}\affiliation{\tunl}\affiliation{\ornl}
\author{C.~Wiseman\,\orcidlink{0000-0002-4232-1326}}\affiliation{\uw}		
\author{W.~Xu}\affiliation{\usd} 
\author{C.-H.~Yu\,\orcidlink{0000-0002-9849-842X}}\affiliation{\ornl}

\collaboration{{\sc{Majorana}} Collaboration}
\noaffiliation

\date{\today}
         
\begin{abstract}
The background index is an important quantity which is used in projecting and calculating the half-life sensitivity of
neutrinoless double-beta decay (\novbb{}) experiments. A novel analysis framework is presented to calculate the background index using the specific activities, masses and simulated efficiencies of an experiment’s components as \textit{distributions}. This Bayesian framework includes a unified approach to combine specific activities from assay. Monte Carlo uncertainty propagation is used to build a background index distribution from the specific activity, mass and efficiency distributions. This analysis method is applied to the {\MJDEMit}, which deployed arrays of high-purity Ge detectors enriched in \Ge{} to search for \novbb{}. The framework projects a mean background index of $\left[8.95 \pm 0.36\right] \times 10^{-4}~\text{cts/(keV\,kg\,yr)}$ from \Th{} and \Ur{} in the {\DEMit}'s components.
\end{abstract}


\maketitle

\section{Introduction} \label{sec:introduction}

A variety of low background experiments form the rich experimental program searching for neutrinoless double-beta decay (\novbb{}). While the decay remains unobserved in all candidate isotopes, the half-life has been constrained to be above $10^{25}-10^{26}$ years by recent experiments~\cite{mjd_0vbb, gerda, kamland_zen, exo, cuore}. The next generation of proposed experiments target a half-life sensitivity, $T_{1/2}$, of $10^{27}-10^{28}$ years~\cite{legend,  nexo, cupid}. To achieve these sensitivities, \novbb{} experiments require underground locations, large isotopic mass and low-background construction materials. Extensive assay screenings are conducted to determine if the experiments structural components meet the targeted background levels. 

In terms of the total electron kinetic energy, the experimental signature of \novbb{} is a peak at the Q-value of the decay, \qval{}. If no background is present in this region of interest (ROI), the sensitivity of a \novbb{} experiment scales linearly with the product of its isotopic mass, $M$, and exposure time, $t$~\cite{sensitivity_bi}. However, if the specific background rate, $b$, is large enough (such that the uncertainty on the background level is proportional to $\sqrt{b \Delta E}$~\cite{review}) the half-life sensitivity ($T_{1/2}$) scales with:
\begin{equation} \label{eq:sensitivity}
T_{1/2} \propto \sqrt{\frac{Mt}{b \Delta E}}.
\end{equation}
The width of the ROI, $\Delta E$, is related to $b$ and the energy resolution at \qval{}. The specific background rate is measured in counts per keV in the ROI, per kg of detector mass, per year (cts/(keV\,kg\,yr)). This observable is also referred to as the background index (BI). Given the low-background nature of \novbb{} experiments, the number of background counts in the ROI is typically too low to estimate the BI (note that no signal would have to be assumed). In such cases, a wider proxy region is needed to increase statistics. This background estimation window (BEW) is often asymmetric around \qval{} to avoid running into the \twovbb{} spectrum and known gamma lines.

Germanium detector technology has been developed for decades, finding applications in radiometric assays and $\gamma$-ray spectroscopy. Ge detectors offer superb energy resolution -- resulting in a narrower ROI -- and can be readily enriched to $\sim 90$\% \Ge{}~\cite{mjd_enr}. The {\MJDEMit} and GERDA experiments exploited this technology to search for \novbb{} in \Ge-enriched high purity Ge (HPGe) detectors~\cite{bb_ge}. Both experiments have the lowest backgrounds of any experiment in the present \novbb{} experimental landscape~\cite{gerda,mjd_0vbb}. 

At the 2039\,keV Q-value of \Ge{} \cite{q_val_1, q_val_2}, the {\DEMit} achieved a BI of $6.2^{+0.6}_{-0.5} \times 10^{-3}$\,cts/(keV\,kg\,yr) in its low background configuration~\cite{mjd_0vbb}. This is not in agreement with the originally projected BI of $<8.75 \times 10^{-4}$\,cts/(keV\,kg\,yr)~\cite{mjd_assay}. The projection was based on simulations of the design geometry and the results of an extensive radioassay program. To account for any discrepancies between the design and as-built geometries of the {\DEMit}, a new set of simulations was performed with the updated geometries. This as-built model did not find a significant deviation from the originally predicted BI. Nevertheless, neither calculation captured the often significant uncertainties from assay in the contribution of natural radiation ($^\text{assay}$BI) to the total BI. Additionally, a systematic review of assay results highlighted the prevalence of measured values that did not agree within error, motivating the development of a technique which properly accounts for this spread and propagates it into the BI.

The total BI includes subdominant contributions from, amongst others, external and cosmogenically induced backgrounds. For this work, only backgrounds quantified by assay -- \Th{}, \Ur{} -- are considered. While the components in the {\DEMit} are also screened for $^{40}$K, it is excluded from the analysis since the $\gamma$-rays produced in the $^{40}$K decay chain have energies well below the BEW.

The $^\text{assay}$BI is calculated by weighting the activity of a component with the {\DEMit}'s simulated array-detection efficiency of decays originating from it. An upper limit on the activity of a component translates into an upper limit on its BI. In the previous projections, such upper limits were directly summed to BIs calculated from measured activities when summing over all modeled components of the {\DEMit}. Thus, the total projected BI was reported as an upper limit itself. Null efficiencies were computed for some components far away from the detector array and failed to contribute to the BI. However, given the high activity of some of these components, their true contribution could still be significant. Therefore, simulation statistics must be properly taken into account.

Other \novbb{} experiments use techniques which address these issues in part. By repeatedly drawing samples from an activity probability density function (PDF) and weighting them by the simulated efficiency, a BI distribution is generated. The CUORE Collaboration generates activity PDFs from fits to preliminary data~\cite{cuore_bgm}. On the other hand, the nEXO Collaboration promotes assay-based activities to a truncated-at-zero Gaussian PDF~\cite{nexo_bgm}. The latter technique is adopted by this work and expanded on by promoting the single-valued efficiency to a distribution as well. Ref.~\cite{assaydistributions} provides a comprehensive summary of the methods used to estimate the BI of various \novbb{} experiments. 

In this article, an assay-based Bayesian framework to project the BI of low-background experiments is presented and applied to the {\MJDEMit}. The framework takes as input all the assay and simulation efficiency data with the goal of:
\begin{enumerate}
    \item Combining multiple assay measurements, including upper limits, using a unified averaging method which properly accounts for the spread in data.
    \item Calculating uncertainties in non-Gaussian regimes such as those posed by null simulation efficiencies. 
    \item Combining uncertainties in assay results, component masses and simulation statistics.
    \item Preserving the generality of this method, allowing for its adoption by the low-background community.
\end{enumerate}

\section{The {\MJDEMit}}\label{sec:mjd}

The {\MJDEMit} consisted of two modules (M1, M2)~\cite{mjd_design} where a total of 40.4\,kg of HPGe detectors (27.2 kg enriched to 88\% in \Ge{}) operated in vacuum at the 4850-foot-level (4300 meter water equivalent) of the Sanford Underground Research Facility (SURF)~\cite{surf} in Lead, South Dakota. The ultra-low background and world-leading energy resolution achieved by the {\DEMit} enabled a sensitive \novbb{} decay search, as well as additional searches for physics beyond the Standard Model.

Both vacuum cryostats and all structural components of the detector arrays were machined from ultra-low-background underground electroformed copper (UGEFCu)~\cite{ugefcu, mjd_assay} and low-background plastics, such as DuPont\textsuperscript{\tiny{TM}} Vespel\textsuperscript{\tiny\textregistered} and polytetrafluoroethylene (PTFE). Low-radioactivity parylene was used to coat UGEFCu threads to prevent galling and for the cryostat seal~\cite{mjd_design}. A layered shield enclosed both modules. The innermost layer consisted of 5\,cm of UGEFCu. Five cm of commercial oxygen-free high conductivity copper (OFHCCu) and 45\,cm of high-purity lead followed. The shield and module volume were constantly purged with low-radon liquid nitrogen boil-off gas. The aluminum enclosure that isolated this Rn-excluded region was covered with a plastic active muon veto which provided near-$4\pi$ coverage~\cite{muonveto}. The near-detector readout system, which was designed for the {\DEMit}, included low-mass front end (LMFE) electronics~\cite{lmfe} and low-mass cables and connectors~\cite{cables}. Cables were guided out of each module following a UGEFCu cross-arm which penetrated the layered shield. The cross-arm connected the cryostat with vacuum and cryogenic hardware. Control and readout electronics were just outside the Rn-excluded region. The entire assembly was surrounded by 5\,cm of borated polyethylene and 25\,cm of pure polyethylene to shield against neutrons.

All components inside the Rn-excluded region populate the as-built model. These were modeled in \textsc{MaGe}, the \textsc{Geant4}-based simulation software jointly developed by the GERDA and {\MJit} collaborations~\cite{mage}. Due to possible radioactive shine through the cross-arm, the vacuum hardware was included as well. The as-built model is based on the Aug. 2016 to Nov. 2019 configuration of the {\DEMit}, where up to 32.1\,kg of detectors were operational. In Nov. 2019, M2 was upgraded with an improved set of cables and connectors and additional cross-arm shielding. The upgrade, in combination with a reconfiguration of M2 detectors, resulted in the final configuration of up to 40.4\,kg of operational HPGe detectors. The final active enriched exposure of the {\DEMit} was 64.5\,kg\,yr. A low-background dataset -- with an active exposure of 63.3\,kg\,yr -- was obtained by excluding data taken previous to the installation of the inner UGEFCu shield. From this dataset, a BI of $6.2^{+0.6}_{-0.5} \times 10^{-3}$\,cts/(keV\,kg\,yr) was calculated~\cite{mjd_0vbb}. The previous data release of the {\DEMit} does not include post-M2-upgrade data. The low-background dataset BI in this release is $4.7\pm0.8 \times 10^{-3}$\,cts/(keV\,kg\,yr) with 21.3\,kg\,yr of active exposure~\cite{mjd_26}. 

To simulate \Th{} and \Ur{} decays originating from the hardware components of the {\DEMit}, their respective decay chains were divided into 10 and 4 segments respectively, following the prescription in Ref.~\cite{gilliss_thesis}. Within each segment secular equilibrium was assumed. The breakpoints in the chains -- which generally correspond to isotopes with half-lives longer than 3 days -- allow a break in secular equilibrium. However, given that the concentration of these isotopes is unknown, secular equilibrium of the chain as a whole was assumed as well. For a particular component and decay chain, the same number of decays were simulated for each segment and the energy depositions in the detectors were recorded. These were later combined with the appropriate branching ratios to produce a spectrum. Section~\ref{sec:bi} describes how the component efficiency is extracted from this simulated spectrum.

\section{A unified approach to combine assay results}\label{sec:assaycomb}

The {\DEMit}'s radioassay program delivered an extensive specific activity database (\Th{}, \Ur{}, $^{40}$K) of the materials used to build the experiment~\cite{mjd_assay}. During the commissioning of the {\DEMit}, additional samples were collected and assayed, thus continually growing this database. Amongst others, inductively coupled plasma mass spectrometry (ICPMS), $\gamma$-count and neutron activation analysis (NAA) measurements were performed. The most sensitive technique, ICPMS, measures specific elements within a decay chain and therefore secular equilibrium is assumed to project a specific activity. The concentration of \Th{} and \Ur{} in HPGe detectors is far too low to be detected by ICPMS measurements. Thus, the \Th{} and \Ur{} contaminations are deduced by searching for time-correlated $\alpha$ decays from their respective decay chains in the {\DEMit}'s low-background dataset. These data-driven results (as opposed to the assay-driven projections published here) will be reported in a future publication. Despite the high detection efficiency of decays originating within Ge, the bulk \Th{} and \Ur{} contamination is anticipated to be so low~\cite{gerda_germanium_assay} that the BI contribution from these sources is expected to be sub-dominant. 

In many cases, measurements of duplicate parts returned specific activities that did not agree within error. A method is  thus needed to properly combine these results. The method should take into account the possible sample-to-sample variation of contaminants and the different detection limits of assay methods. It should also allow for the inclusion of assays which result in upper limits.

Following the methodology of the Particle Data Group (PDG) for unconstrained averaging, a standard weighted least-squares approach is employed~\cite{PDG}. The average specific activity and its uncertainty are calculated as
\begin{equation} \label{eq:unconstrained_averaging}
\overline{a} \pm \delta \overline{a} = \frac{\sum_k w_k a_k}{\sum_k w_k} \pm \left( \sum_k w_k \right)^{-1/2},
\end{equation}
where the $k$-th specific activity from assay, $a_k$, is weighted by 
\begin{equation}
w_k = 1 / (\delta a_k)^2.
\end{equation}
$N_a$ assays populate the sum, including all assays resulting in a measured value. The PDG states ``We do not average or combine upper limits except in a very few cases where they may be re-expressed as measured numbers with Gaussian errors.''~\cite{PDG} Exactly the latter is used to treat specific activity upper limits. More concisely, only the \textit{most stringent} 90\% C.L. upper limit, $l$, is re-expressed  as a measured number, $a_k = 0$, with Gaussian error $\delta a_k = l/[\sqrt{2} \text{erf}^{-1}(0.9)]$. Note that an issue would arise if multiple upper limits are combined. Combining two identical $a_k \pm \delta a_k$ results in a smaller combined uncertainty $\delta \overline{a} = \delta a_k/\sqrt{2}$. This is a desirable property for measured values, but not for upper limits. Repeated results are likely if the activity of a source is significantly below the detection limit of the assay apparatus. Combining such measurements would result in an unrealistically lowered upper limit. Therefore only the most stringent upper limit is chosen. A maximum of one upper limit is thus included in the sum of Eq.~\ref{eq:unconstrained_averaging}.

Once the average is computed, $\chi^2 = \sum_k w_k (\overline{a} - a_k)^2$ is calculated. $\chi^2/(N_a-1)$ is used as a discriminant as follows:
\begin{enumerate}
	\item If $\chi^2 / (N_a-1) \leq 1$ the average, $\overline{a}$, and uncertainty, $\delta \overline{a}$, are accepted.
	\item If $\chi^2 / (N_a-1) > 1$ the uncertainty, $\delta \overline{a}$, is scaled by a factor $\Sigma = \left[ \chi^2/ (N_a-1) \right]^{1/2}$.
\end{enumerate}
\begin{figure}[htbp]
    \centering
    \includegraphics[width=1\linewidth]{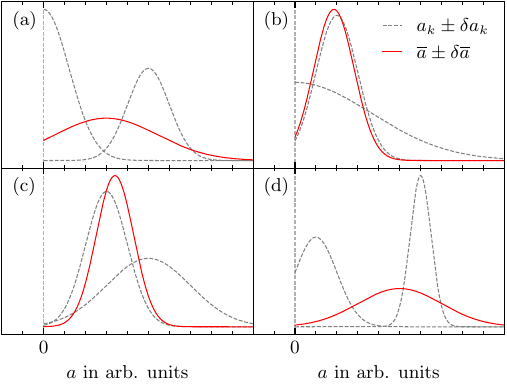}
    \caption{The combination of an upper limit with a measured value (a,b) and the combination of two measured values (c,d) are shown as truncated-at-zero Gaussian distributions in solid red. Toy data is used for the raw upper limits and measured values, which are also depicted as truncated-at-zero Gaussian distributions but in dashed grey.}
    \label{fig:combined_assays}
\end{figure}
Fig.~\ref{fig:combined_assays} exemplifies the technique for combining assay results. Following the nEXO collaboration's treatment of activities, these are visualized as truncated-at-zero Gaussian distributions. In the figure two example assay results, $a_k\,\pm\,\delta a_k$, are plotted as such. These are averaged using the technique described above and the result, $\overline{a}\,\pm\,\delta\overline{a}$, is shown. An upper limit lower than a measured value lowers the measured value~Fig.~\ref{fig:combined_assays}(a), whereas an upper limit higher than a measured value has almost no effect on the same~Fig.~\ref{fig:combined_assays}(b). Note that in Fig.~\ref{fig:combined_assays}(a) a scaling factor of $\Sigma$ is applied since $\chi^2>1$. If two measured values agree within error, the combined uncertainty decreases~Fig.~\ref{fig:combined_assays}(c). If there is no agreement, the combined uncertainty increases since a scaling factor of $\Sigma$ is applied~Fig.~\ref{fig:combined_assays}(d).

\begin{figure}[htbp]
    \centering
    \includegraphics[width=1\linewidth]{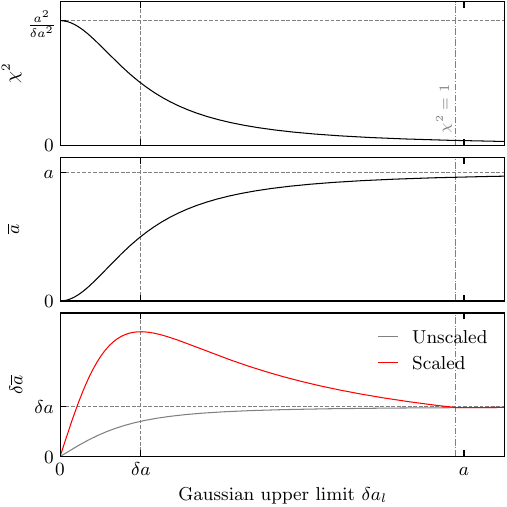}
    \caption{The $\chi^2$, average ($\overline{a}$) and uncertainty ($\delta\overline{a}$), of a measured value, $a \pm \delta a$, and the Gaussian upper limit, $\delta a_l$, defined in the text. $\delta a$ is set to $0.2a$. A dashed and dash-dotted vertical line is shown at $\delta a_l = \delta a$ and at the $\delta a_l$ which results in $\chi^2 = 1$.}
    \label{fig:including_ul}
\end{figure}

The levels of \Th{} and \Ur{} in the {\DEMit}'s components often lie below the detection limits of the most sensitive assays, leading to a high prevalence of upper limits. The motivation to include upper limits, not only in the sum of Eq.~\ref{eq:unconstrained_averaging} but also in the calculation of the scaling factor $\Sigma$, stems from this prevalence. To justify their use, the effect of combining an upper limit, $l$, with a fixed measured value, $a\,\pm\,\delta a$, with small error ($\delta a < a$), was evaluated. This case is representative of many assay results. Fig.~\ref{fig:including_ul} shows the $\chi^2$, average and uncertainty with respect to a Gaussian upper limit, $\delta a_l$, between 0 and $1.1a$. It demonstrates that averaging with an upper limit leads to the following desirable properties:
\begin{enumerate}
	\item For $\delta a_l > a + \delta a$, $\overline{a}$ and $\delta\overline{a}$ rapidly converge to $a$ and $\delta a$ respectively. Smaller $\delta a$ lead to faster convergence. In other words, upper limits higher than measured values have little effect on the same. The lower the measured value's error, the lower the impact. In this regime $\chi^2 < 1$.
	\item For $\delta a_l < a + \delta a$, $\overline{a}$ decreases monotonically with $\delta a_l$.
        \item The uncertainty, $\delta\overline{a}$ is maximized when $\delta a_l = \delta a$. For this value, $\overline{a} = a/2$. Note the importance of scaling the uncertainty and thus the need to include the upper limit in the calculation of $\Sigma$.
\end{enumerate}
If assays have to be averaged more than once, multiple upper limits may contribute to the result. For example, 40 of the {\DEMit}'s lead bricks were assayed with the following procedure. One or more samples were taken from each brick, and each sample was used to prepare one or more dilutions to conduct ICPMS measurements. In some cases each dilution was separated into multiple vials, which were separately measured. Thus averaging is performed at the dilution, sample, brick and global levels, with upper limits contributing at each stage. 

The specific activity averages and uncertainties of the materials used in the {\MJDEMit} are presented in Table~\ref{tab:assay}. These values are used to calculate the BI contributions of all \Th{} and \Ur{} sources used to model the experiment. In this table, $N_a$ represents the number of assays contributing at the global averaging level. While the set of assayed components is limited, it is assumed to be representative of all those used in the {\DEMit}.
\begin{table*}[htbp]
    \caption{Measured ($N_a = 1$) or combined ($N_a > 1$, using the technique described in Section~\ref{sec:assaycomb}) bulk \Th{} and \Ur{} specific activities of materials used in the {\MJDEMit}. Where applicable, the most stringent 90\% C.L. upper limit is used. The assay methodology is the same for all $N_a$ assays in a given row. If $N_a$ is different for \Th{} and \Ur{}, then comma separated values are given. Machined UGEFCu samples were fabricated from stock UGEFCu. The latter is assumed for the specific activity of the inner copper shield of the {\DEMit}. Signal cables were separated into two sections. Custom connectors were designed and built to connect these sections, with the female end wired to an LMFE and the male end wired to pre-amplification and readout electronics. When calculating the BI, all components made from a given material are collected in the group shown in the second column. A total of $N_a$ assays -- sourced from Ref.~\cite{mjd_assay} and additional measurement campaigns -- are combined. }
    \label{tab:assay}
    \begin{ruledtabular}
    \begin{tabular}{l c c c c c}
    	Material & Group & \Th{} [$\upmu$Bq/kg] & \Ur{} [$\upmu$Bq/kg] & $N_a$ & Method\\
    	\hline\noalign{\vskip 1ex}
    	LMFE       & Front Ends & 6950 $\pm$ 830             & 10600 $\pm$ 300 & 2 & ICPMS\\
    	HV Cable    & Cables & 87.9 $\pm$ 52.4             & 231 $\pm$ 34 & 3  & ICPMS\\
    	Signal Cable   & Cables & 546 $\pm$ 112             & 530 $\pm$ 58 & 3  & ICPMS\\
    	Stock UGEFCu   & Electroformed Cu & 0.188 $\pm$ 0.029         & 0.137 $\pm$  0.044 & 5 & ICPMS\\
    	Machined UGEFCu  & Electroformed Cu & 0.575 $\pm$ 0.088             & 0.752 $\pm$  0.083 & 12 &  ICPMS\\
    	OFHCCu    & OFHC Cu Shielding &  1.10 $\pm$ 0.14            & 1.37 $\pm$  0.18 & 2 & ICPMS\\
    	Pb Bricks  &  Pb Shielding     & 9.53 $\pm$  1.01             & 25.6 $\pm$ 1.5 & 29, 19 & ICPMS\\
    	Female Connector  & Connectors & 390 $\pm$ 7            & 540 $\pm$  9 & 1 & ICPMS\\
    	Male Connector  & Connectors & 28.8 $\pm$ 2.0              & 130 $\pm$  11 & 1 & ICPMS\\
    	PTFE O-ring  & Other Plastics & 39.7 $\pm$ 31.9             & $<$ 105 & 2, 1 & NAA\\
    	PTFE       & Detector Unit PTFE & 0.101 $\pm$ 0.008              & $<$4.97 & 1 & NAA\\
    	DuPont\textsuperscript{\tiny{TM}} Vespel\textsuperscript{\tiny\textregistered}     & Other Plastics &  360 $\pm$ 234             & 403 $\pm$ 179 & 3 & ICPMS\\
    	PTFE Gasket & Other Plastics & $<$20.7              & $<$94.5 & 1 & NAA\\
    	Stainless Steel & Vacuum Hardware &  13000 $\pm$  4000            & $<$5000  & 1 & $\gamma$-count\\
    	Glass Break & Vacuum Hardware & 49000 $\pm$ 8000              & 160000 $\pm$ 10000 & 1 & $\gamma$-count\\
    	PTFE Tubing & Other Plastics & 6.09 $\pm$ 7.30 &  $<$38.6 & 1 & ICPMS\\
    	Parylene & Parylene & 2150 $\pm$ 120 & 3110 $\pm$ 750 & 1 & ICPMS\\
	\end{tabular}
        \end{ruledtabular}
\end{table*}

\section{Background index}\label{sec:bi}

Simulations predict an approximately flat background in the 370\,keV split BEW covering 1950-2350\,keV. The BEW has three 10\,keV regions removed, centered at 2103 keV, the $^{208}$Tl (\Th{}) single escape peak, and 2118\,keV and 2204\,keV, the $^{214}$Bi (\Ur{}) peaks. This BEW is used to estimate the BI at \qval{}, for both simulations and data. When estimating the BI from data an additional 10\,keV region, centered at \qval{}, is removed~\cite{mjd_0vbb}. 

The contribution to the {\DEMit}'s BI from a natural radiation source $i$ -- the \Th{} or \Ur{} contamination in a component of mass $m_i$ -- is calculated as follows. For each segment, $j$, of the decay chain of the source's contaminant, \MaGe{} simulates $N_{ij}$ decays. For a given source $i$, all $N_{ij}$ are equal up to computational errors. Simulated events leading to energy depositions in an operational detector are subject to the same anti-coincidence and pulse shape analysis cuts applied to data (designed to select \novbb{}-like single-site events). The number of counts passing all cuts, $n_{ij}$, is calculated by integrating the resulting combined detector spectra over the BEW. The corresponding segment efficiency is thus:
\begin{equation} \label{eq:eff}
    \epsilon_{ij}^{\textsc{Mjd}}  = n_{ij}/N_{ij}
\end{equation}
The segment efficiencies are weighted by the branching ratio of each segment, $\mathcal{B}_j$, and summed over the total number of segments, $S_i$, to produce the source efficiency, $\epsilon_i^{\textsc{Mjd}}$:
\begin{equation} \label{eq:eff_tot}
    \epsilon_i^{\textsc{Mjd}} =  \sum_j^{S_i} ~\mathcal{B}_j \epsilon_{ij}^{\textsc{Mjd}}
\end{equation}
The source efficiency, mass and averaged specific activity, $\overline{a}_i$, are used to calculate the BI of the source:
\begin{equation} \label{eq:bi}
    \text{BI}_i = \frac{1}{M \times \Delta} \epsilon_i^{\textsc{Mjd}} \times \overline{a}_i \times m_i
\end{equation}
Where $\Delta = 370$\,keV is the width of the BEW and $M$ is the mass of operational detectors in the array. Note that the total number of sources is equal to twice the number of components used to model the {\DEMit}, since \Th{} and \Ur{} are accounted for in all components. The source contributions to the total $^\text{assay}\text{BI} = \sum_i \text{BI}_i$, can be summed as desired, either by source material, contaminant, or component group. If only an upper limit, $l_i$, is available, it can be used in place of $\overline{a}_i$. This calculates a BI which is the direct sum of upper limits and central values and is referred to as the direct BI in the text. The component-group-combined \Th{} and \Ur{} direct BIs are calculated by summing over the appropriate sources and are presented in Table~\ref{tab:bi}. Note that the design geometry projection of Ref.~\cite{mjd_assay} used this method. It is not possible to assign an uncertainty to the direct BI because of the inclusion of upper limits. However, a proper treatment of uncertainties can be obtained by promoting the single-valued $\overline{a}_i$ (or $l_i$), $m_i$ and $\epsilon^{\textsc{Mjd}}_i$ to distributions and using Monte Carlo uncertainty propagation to combine these in a final BI distribution. 

\section{Promoting single values to distributions}\label{sec:distributions}

As described in Section~\ref{sec:assaycomb} the specific activity can be expressed as a truncated-at-zero Gaussian. The mass, $m_i$, is also promoted to a distribution of this form. Depending on the method used to measure or estimate the mass, a Gaussian uncertainty is assigned. This uncertainty ranges from 1\%, to account for sample to sample variation in direct mass measurements, to 10\% for masses that were estimated from geometry. In practice the mass PDFs are indistinguishable from a true Gaussian given the assigned level of uncertainty.

The derivation of the efficiency, $\epsilon^{\textsc{Mjd}}_i$, PDF follows. Starting from Eq.~\ref{eq:eff}, the probability of finding $n$ counts in the BEW -- with expectation value $\lambda$ -- is given by the Poisson distribution, $P(n|\lambda) = \lambda^ne^{-\lambda}/n!$. Conversely, $P(\lambda|n)$ is deduced via Bayes' theorem, $P(\lambda|n) \propto P(\lambda)P(n|\lambda)$, using the prior, 
\begin{equation}
    P(\lambda) = \frac{1}{\lambda^{1-1/S}}.
\end{equation}
The choice of uninformative prior corresponds to a flat prior on the sum of the $S$ decay chain segments. The functional form of the prior is derived in Appendix~\ref{app:prior}.
\begin{align} \label{eq:gamma}
    P(\lambda|n) &\propto \frac{\lambda^{n-1+1/S}e^{-\lambda}}{n!} \nonumber\\
    &\propto  \Gamma(\lambda| \alpha = n + 1/S,~\beta = 1)
\end{align}
The resulting posterior is a Gamma distribution with parameters $\alpha = n + 1/S$ and $\beta = 1$. The probability of obtaining an efficiency $\epsilon_i^{\textsc{Mjd}}$, given an ensemble of counts $\{n_{ij}\}$ is thus,
\begin{align} \label{eq:eff_tot_pdf}
    P(\epsilon_i^{\textsc{Mjd}}|\{n_{ij}\}) &= \sum_j^{S_i} ~\mathcal{B}_j P(\epsilon_{ij}^{\textsc{Mjd}}|n_{ij})  \nonumber\\
    &=  \sum_j^{S_i} ~\mathcal{B}_j \frac{P(\lambda_{ij}|n_{ij})}{N_{ij}}~.
\end{align}
This result is obtained by replacing $n_{ij}$ by $P(\lambda_{ij}|n_{ij})$ in Eq.~\ref{eq:eff} and carrying it into Eq.~\ref{eq:eff_tot}. The PDF of $\epsilon_i^{\textsc{Mjd}}$ is computed numerically. Taking a random draw from the PDF in Eq.~\ref{eq:gamma} for each segment, weighing by the appropriate factors and summing them as in Eq.~\ref{eq:eff_tot_pdf}. This process is repeated $>10^5$ times to generate the PDF or only once if one efficiency sample is needed. 

Not all segments are included in the sum for far away sources, only segments which produce $\gamma$'s with sufficiently high energies to lead to energy depositions in the BEW. The excluded segments have a true zero efficiency.

\section{Monte Carlo uncertainty propagation}\label{sec:mcup}

The assay-based background index can be promoted to a distribution by sampling the PDFs described in Section~\ref{sec:distributions}. The $\text{BI}_i$ distributions are thus generated in a similar manner as that of the efficiency. A random draw is taken from the specific activity, mass, and efficiency PDFs. These are multiplied and scaled as in Eq.~\ref{eq:bi}. The process is repeated $10^6$ times to produce a PDF for the BI of the source.
Each set of draws constitutes a toy experiment resulting in a different BI. The uncertainty of the efficiency, specific activity and mass -- embedded in their corresponding PDFs -- is propagated into the $\text{BI}_i$ distribution through this process, referred to as Monte Carlo uncertainty propagation. 

\begin{figure}[htb]
    \centering
    \includegraphics[width=1\linewidth]{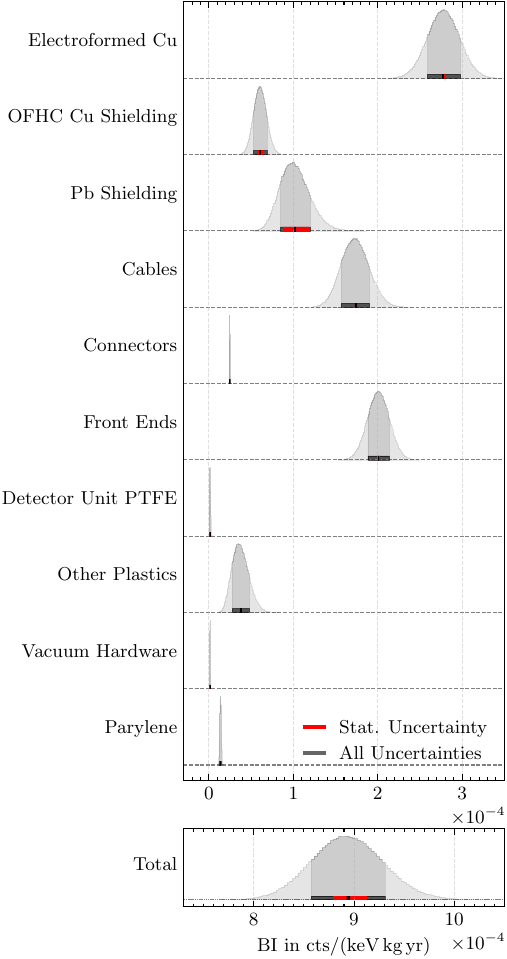}
    \caption{The mean BI and 1$\sigma$ uncertainties of the {\DEMit}'s component groups and their sum are plotted as a vertical black lines through dark grey error bars. Statistical uncertainty contributions are shown in red. These are overlaid on their corresponding PDFs in light grey (with the central 2$\sigma$ region shaded). All PDFs contain $10^6$ samples and are scaled vertically such that they have the same height.}
    \label{fig:backgroundbudget}
\end{figure}

The $\text{BI}_i$ PDFs are combined by taking the direct sum of the samples that were used to generate them. Fig.~\ref{fig:backgroundbudget} shows the result of summing all the $\text{BI}_i$ PDFs belonging to a component group in the {\DEMit}. The mean and uncertainties that are extracted from these distributions are reported in Table~\ref{tab:bi} and plotted as error bars in Fig.~\ref{fig:backgroundbudget}. The total BI distribution of \Th{} and \Ur{} sources in the {\DEMit} is computed in a similar manner and displayed at the bottom of the figure. 

Simulation statistical uncertainties can be isolated by collapsing the specific activity and mass distributions to their respective means and using them to weigh the samples that were drawn from the efficiency distribution. This information can be used by the analyst as a guide to set the number of decays in future simulations.

Monte Carlo uncertainty propagation allows for all potential sources of natural radiation to be incorporated in the model, not just those which produced counts in the BEW. With no counts in the BEW, the single-valued efficiency is zero. However, the efficiency \textit{distribution} will take an exponential form, effectively setting an upper limit on the BI (once weighted properly) for the source in question. This upper limit is fully dependent on the number of simulated decays and given enough computational resources should be driven down to the point where it is negligible compared to other contributions to the BI. This is the case for the Vacuum Hardware and Pb Shielding \Ur{} contamination.
\begin{table*}[htb]
    \caption{Background indices and $1\sigma$ uncertainties of the component groups of the {\DEMit} as derived from their distributions. A 90\% C.L. upper limit is given if the width of the central $2\sigma$ region is higher than the width of the lower $2\sigma$ region. The direct BIs are included for reference.}
    \label{tab:bi}
    \begin{ruledtabular}
    \begin{tabular}{l c c c c}
    	Group & Mean \Th{} BI & Mean \Ur{} BI & Mean BI & Direct BI\\
    	& [cts/(keV\,kg\,yr)] & [cts/(keV\,kg\,yr)] & [cts/(keV\,kg\,yr)] & [cts/(keV\,kg\,yr)]\\
        \hline\noalign{\vskip 1ex}
        \quad Electroformed Cu & $2.44^{+0.19}_{-0.19} \times 10^{-4}$ & $3.41^{+0.23}_{-0.23} \times 10^{-5}$ & $2.78^{+0.19}_{-0.19} \times 10^{-4}$ & $2.75 \times 10^{-4}$ \\ \noalign{\vskip 0.5ex}
        \quad OFHC Cu Shielding & $5.49^{+0.77}_{-0.77} \times 10^{-5}$ & $5.79^{+1.21}_{-1.21} \times 10^{-6}$ & $6.07^{+0.78}_{-0.78} \times 10^{-5}$ & $6.05 \times 10^{-5}$ \\ \noalign{\vskip 0.5ex}
        \quad Pb Shielding & $7.75^{+1.38}_{-1.38} \times 10^{-5}$ & $2.52^{+1.06}_{-1.06} \times 10^{-5}$ & $1.03^{+0.18}_{-0.18} \times 10^{-4}$ & $9.70 \times 10^{-5}$ \\ \noalign{\vskip 0.5ex}
        \quad Cables & $1.41^{+0.17}_{-0.17} \times 10^{-4}$ & $3.29^{+0.20}_{-0.20} \times 10^{-5}$ & $1.74^{+0.17}_{-0.17} \times 10^{-4}$ & $1.93 \times 10^{-4}$ \\ \noalign{\vskip 0.5ex}
        \quad Connectors & $2.09^{+0.03}_{-0.03} \times 10^{-5}$ & $3.46^{+0.07}_{-0.07} \times 10^{-6}$ & $2.44^{+0.03}_{-0.03} \times 10^{-5}$ & $2.44 \times 10^{-5}$ \\ \noalign{\vskip 0.5ex}
        \quad Front Ends & $1.54^{+0.13}_{-0.13} \times 10^{-4}$ & $4.67^{+0.10}_{-0.10} \times 10^{-5}$ & $2.01^{+0.13}_{-0.13} \times 10^{-4}$ & $2.15 \times 10^{-4}$ \\ \noalign{\vskip 0.5ex}
        \quad Detector Unit PTFE & $1.96^{+0.18}_{-0.18} \times 10^{-7}$ & $1.05^{+0.57}_{-0.57} \times 10^{-6}$ & $1.24^{+0.57}_{-0.57} \times 10^{-6}$ & $2.36 \times 10^{-6}$ \\ \noalign{\vskip 0.5ex}
        \quad Other Plastics & $3.18^{+0.97}_{-0.97} \times 10^{-5}$ & $5.99^{+2.53}_{-2.52} \times 10^{-6}$ & $3.78^{+1.00}_{-1.00} \times 10^{-5}$ & $4.10 \times 10^{-5}$ \\ \noalign{\vskip 0.5ex}
        \quad Vacuum Hardware & $1.06^{+0.25}_{-0.25} \times 10^{-6}$ & $<$ $7.80 \times 10^{-8}$ & $1.10^{+0.25}_{-0.25} \times 10^{-6}$ & $9.41 \times 10^{-7}$ \\
        \quad Parylene & $1.10^{+0.08}_{-0.08} \times 10^{-5}$ & $2.55^{+0.48}_{-0.48} \times 10^{-6}$ & $1.35^{+0.09}_{-0.09} \times 10^{-5}$ & $1.35 \times 10^{-5}$ \\ \noalign{\vskip 1ex}
        \hline\noalign{\vskip 1ex}
        SUM & $7.37^{+0.34}_{-0.34} \times 10^{-4}$ & $1.58^{+0.11}_{-0.11} \times 10^{-4}$ & $8.95^{+0.36}_{-0.36} \times 10^{-4}$ & $9.23 \times 10^{-4}$ \\
    \end{tabular}
    \end{ruledtabular}
\end{table*}

\section{Projected background and conclusions}\label{sec:results}

As Section \ref{sec:mcup} describes, the mean BI and the corresponding uncertainties of the {\MJDEMit}'s component groups are extracted from their distributions in Fig.~\ref{fig:backgroundbudget} and reported in Table~\ref{tab:bi}. The mean total natural radiation BI, determined from its distribution at the bottom of the figure, is
\begin{equation*} \label{eq:bi_nat}
    ^{\text{assay}}\text{BI} = \left[8.95 \pm 0.16 (\text{stat.}) \pm 0.20 (\text{act.})\right] \times 10^{-4}
\end{equation*}
in units of cts/(keV\,kg\,yr). The uncertainty from activity (act.) dominates $^{\text{assay}}\text{BI}$ and most component group BIs. The statistical uncertainty from simulation (stat.) has been calculated with the method described in Section~\ref{sec:mcup}. Note that the symmetric uncertainty around the mean does not capture the asymmetry of the distribution. The contributions from the \Th{} and \Ur{} decay chains, given in units of cts/(keV\,kg\,yr), follow:
\begin{align*}
    ^{\text{232}}\text{BI} &= \left[7.37 \pm 0.12 (\text{stat.}) \pm 0.22 (\text{act.})\right] \times 10^{-4}~,\\
    ^{\text{238}}\text{BI} &= \left[1.58 \pm 0.11 (\text{stat.}) \pm 0.01 (\text{act.})\right] \times 10^{-4}~.
\end{align*}

By comparing the direct BI of Table~\ref{tab:bi} with the design geometry projection of Ref.~\cite{mjd_assay} for the same components, a 44\% increase is found. Uncertainties were not calculated for the design geometry projection. On the other hand, Monte Carlo uncertainty propagation captured the uncertainties from specific activity, component mass and simulation efficiencies in the BI. Multiple contaminants, which have null single-valued efficiencies, contribute to the BI under this new framework. 

The projected $^{\text{assay}}\text{BI}$ does not capture the uncertainty associated with the assumption of secular equilibrium or account for the possible introduction of backgrounds during the construction of the {\DEMit}. Additionally, given the limited number of assayed components, the variation in contaminants may be larger than captured by the averaged assay uncertainty. Furthermore, possible systematic uncertainties in simulated component geometry were not taken into account. Nevertheless, the Monte Carlo uncertainty propagation framework that has been developed can be extended to account for these effects.
 
The techniques outlined in this article can be exploited to project the BI of future \novbb{} experiments. Particularly, the design of such experiments can benefit from the ability to include component mass uncertainties, since the designed and as-built geometries often differ. Additionally, the framework informs the analyst on the number of decays to simulate, thus optimizing computational resources. The unified approach to average assay results can facilitate the standardization of assay reporting. This is of importance in the field, given that the design of new experiments often draws from assay data collected by others. The uncertainty extracted from the BI can be propagated into the projected sensitivity, thus further illuminating the physics reach of the next generation of \novbb{} experiments.

\section*{Acknowledgments}
This material is based upon work supported by the U.S.~Department of Energy, Office of Science, Office of Nuclear Physics under contract / award numbers DE-AC02-05CH11231, DE-AC05-00OR22725, DE-AC05-76RL0130, DE-FG02-97ER41020, DE-FG02-97ER41033, DE-FG02-97ER41041, DE-SC0012612, DE-SC0014445, DE-SC0017594, DE-SC0018060, DE-SC0022339, and LANLEM77/LANLEM78. We acknowledge support from the Particle Astrophysics Program and Nuclear Physics Program of the National Science Foundation through grant numbers MRI-0923142, PHY-1003399, PHY-1102292, PHY-1206314, PHY-1614611, PHY-1812409, PHY-1812356, PHY-2111140, and PHY-2209530. We gratefully acknowledge the support of the Laboratory Directed Research \& Development (LDRD) program at Lawrence Berkeley National Laboratory for this work. We gratefully acknowledge the support of the U.S.~Department of Energy through the Los Alamos National Laboratory LDRD Program, the Oak Ridge National Laboratory LDRD Program, and the Pacific Northwest National Laboratory LDRD Program for this work.  We gratefully acknowledge the support of the South Dakota Board of Regents Competitive Research Grant. 
We acknowledge the support of the Natural Sciences and Engineering Research Council of Canada, funding reference number SAPIN-2017-00023, and from the Canada Foundation for Innovation John R.~Evans Leaders Fund.  
We acknowledge support from the 2020/2021 L'Or\'eal-UNESCO for Women in Science Programme.
This research used resources provided by the Oak Ridge Leadership Computing Facility at Oak Ridge National Laboratory and by the National Energy Research Scientific Computing Center, a U.S.~Department of Energy Office of Science User Facility. We thank our hosts and colleagues at the Sanford Underground Research Facility for their support.

\appendix
\section{Efficiency distribution prior derivation}\label{app:prior}

The choice of prior, $P(\lambda_{ij}) = 1/\lambda_ij^{1-1/S_i}$, becomes apparent here when all $n_{ij}$, $B_{j}$, $N_{ij}$ are equal for a given $i$ in Eq.~\ref{eq:eff_tot}. Setting $n_{ij} = n_j$ and dropping the $i$ indices, the sum to be evaluated is $\sum_j^S P(\lambda_j|n)$. The sum of PDFs is given by their convolution. Defining the convolution $f$ with itself $S-1$ times as
\begin{equation}
    f^{\{S\}} = \underbrace{f * (f * (f * \cdots * (f * f)))}_{S-1\text{ convolutions}},
\end{equation}
and rewriting $P(\lambda_j|n)$ as $f(\lambda_j)$, an unknown function to be solved for, the sum becomes the following convolution:
\begin{equation}
    \sum_j^S~f(\lambda_j) = f^{\{S\}}(\lambda_j)
\end{equation}
If the decay chain was simulated as a whole and not by segment, $P(\lambda|n)$ would be given by Eq.~\ref{eq:gamma} but with $S = 1$ and $n = Sn_j$. The functional form of $f(\lambda_j)$ must be such that the sum of the PDFs of the segments equals the PDF of the full decay chain:
\begin{equation}
    f^{\{S\}}(\lambda_j)\ = \lambda^{Sn_j}e^{-\lambda}
\end{equation}
Taking the Laplace transform and working simultaneously on both sides yields:
\begin{align}\label{eq:eff_seg}
   \mathscr{L}\left[f^{\{S\}}(\lambda_j)\right](t) &= \mathscr{L}\left[\lambda^{Sn_j}e^{-\lambda}\right](t) \nonumber\\
    \Bigl\{ \mathscr{L}\left[f(\lambda_j)\right](t) \Bigl\}^S &= \frac{(Sn_j)!}{(t+1)^{Sn_j+1}} \nonumber\\
    f(\lambda_j) &=  \sqrt[S]{(Sn_j)!}~\mathscr{L}^{-1}\left[\frac{1}{(t+1)^{n_j+1/S}}\right](\lambda_j) \nonumber\\
    &=  \sqrt[S]{(Sn_j)!}~\frac{\lambda_j^{n_j-1+1/S}e^{-\lambda_j}}{\Gamma(n_j + 1/S)}
\end{align}
Eq.~\ref{eq:eff_seg} has the same functional form as Eq.~\ref{eq:gamma} up to a constant. Therefore 
when $n_{ij}$, $B_{j}$, $N_{ij}$ are equal for a given $i$, the prior must take the form $P(\lambda_{j}) = 1/\lambda_j^{1-1/S}$. The use of this prior is extended to all cases.

\providecommand{\noopsort}[1]{}\providecommand{\singleletter}[1]{#1}%


\begin{thebibliography}{29}%
\makeatletter
\providecommand \@ifxundefined [1]{%
 \@ifx{#1\undefined}
}%
\providecommand \@ifnum [1]{%
 \ifnum #1\expandafter \@firstoftwo
 \else \expandafter \@secondoftwo
 \fi
}%
\providecommand \@ifx [1]{%
 \ifx #1\expandafter \@firstoftwo
 \else \expandafter \@secondoftwo
 \fi
}%
\providecommand \natexlab [1]{#1}%
\providecommand \enquote  [1]{``#1''}%
\providecommand \bibnamefont  [1]{#1}%
\providecommand \bibfnamefont [1]{#1}%
\providecommand \citenamefont [1]{#1}%
\providecommand \href@noop [0]{\@secondoftwo}%
\providecommand \href [0]{\begingroup \@sanitize@url \@href}%
\providecommand \@href[1]{\@@startlink{#1}\@@href}%
\providecommand \@@href[1]{\endgroup#1\@@endlink}%
\providecommand \@sanitize@url [0]{\catcode `\\12\catcode `\$12\catcode `\&12\catcode `\#12\catcode `\^12\catcode `\_12\catcode `\%12\relax}%
\providecommand \@@startlink[1]{}%
\providecommand \@@endlink[0]{}%
\providecommand \url  [0]{\begingroup\@sanitize@url \@url }%
\providecommand \@url [1]{\endgroup\@href {#1}{\urlprefix }}%
\providecommand \urlprefix  [0]{URL }%
\providecommand \Eprint [0]{\href }%
\providecommand \doibase [0]{https://doi.org/}%
\providecommand \selectlanguage [0]{\@gobble}%
\providecommand \bibinfo  [0]{\@secondoftwo}%
\providecommand \bibfield  [0]{\@secondoftwo}%
\providecommand \translation [1]{[#1]}%
\providecommand \BibitemOpen [0]{}%
\providecommand \bibitemStop [0]{}%
\providecommand \bibitemNoStop [0]{.\EOS\space}%
\providecommand \EOS [0]{\spacefactor3000\relax}%
\providecommand \BibitemShut  [1]{\csname bibitem#1\endcsname}%
\let\auto@bib@innerbib\@empty
%</preamble>
\bibitem [{\citenamefont {Arnquist}\ \emph {et~al.}(2023)\citenamefont {Arnquist} \emph {et~al.}}]{mjd_0vbb}%
  \BibitemOpen
  \bibfield  {author} {\bibinfo {author} {\bibfnamefont {I.~J.}\ \bibnamefont {Arnquist}} \emph {et~al.} (\bibinfo {collaboration} {\textsc{Majorana} Collaboration}),\ }\href {https://doi.org/10.1103/PhysRevLett.130.062501} {\bibfield  {journal} {\bibinfo  {journal} {Phys. Rev. Lett.}\ }\textbf {\bibinfo {volume} {130}},\ \bibinfo {pages} {062501} (\bibinfo {year} {2023})}\BibitemShut {NoStop}%
\bibitem [{\citenamefont {Agostini}\ \emph {et~al.}(2020)\citenamefont {Agostini} \emph {et~al.}}]{gerda}%
  \BibitemOpen
  \bibfield  {author} {\bibinfo {author} {\bibfnamefont {M.}~\bibnamefont {Agostini}} \emph {et~al.} (\bibinfo {collaboration} {GERDA Collaboration}),\ }\href {https://doi.org/10.1103/PhysRevLett.125.252502} {\bibfield  {journal} {\bibinfo  {journal} {Phys. Rev. Lett.}\ }\textbf {\bibinfo {volume} {125}},\ \bibinfo {pages} {252502} (\bibinfo {year} {2020})}\BibitemShut {NoStop}%
\bibitem [{\citenamefont {Abe}\ \emph {et~al.}(2023)\citenamefont {Abe} \emph {et~al.}}]{kamland_zen}%
  \BibitemOpen
  \bibfield  {author} {\bibinfo {author} {\bibfnamefont {S.}~\bibnamefont {Abe}} \emph {et~al.} (\bibinfo {collaboration} {KamLAND-Zen Collaboration}),\ }\href {https://doi.org/10.1103/PhysRevLett.130.051801} {\bibfield  {journal} {\bibinfo  {journal} {Phys. Rev. Lett.}\ }\textbf {\bibinfo {volume} {130}},\ \bibinfo {pages} {051801} (\bibinfo {year} {2023})}\BibitemShut {NoStop}%
\bibitem [{\citenamefont {Anton}\ \emph {et~al.}(2019)\citenamefont {Anton} \emph {et~al.}}]{exo}%
  \BibitemOpen
  \bibfield  {author} {\bibinfo {author} {\bibfnamefont {G.}~\bibnamefont {Anton}} \emph {et~al.} (\bibinfo {collaboration} {EXO-200 Collaboration}),\ }\href {https://doi.org/10.1103/PhysRevLett.123.161802} {\bibfield  {journal} {\bibinfo  {journal} {Phys. Rev. Lett.}\ }\textbf {\bibinfo {volume} {123}},\ \bibinfo {pages} {161802} (\bibinfo {year} {2019})}\BibitemShut {NoStop}%
\bibitem [{\citenamefont {Adams}\ \emph {et~al.}(2022)\citenamefont {Adams} \emph {et~al.}}]{cuore}%
  \BibitemOpen
  \bibfield  {author} {\bibinfo {author} {\bibfnamefont {D.~Q.}\ \bibnamefont {Adams}} \emph {et~al.} (\bibinfo {collaboration} {CUORE Collaboration}),\ }\href {https://doi.org/10.1038/s41586-022-04497-4} {\bibfield  {journal} {\bibinfo  {journal} {Nature}\ }\textbf {\bibinfo {volume} {604}},\ \bibinfo {pages} {53} (\bibinfo {year} {2022})}\BibitemShut {NoStop}%
\bibitem [{\citenamefont {Abgrall}\ \emph {et~al.}(2017)\citenamefont {Abgrall} \emph {et~al.}}]{legend}%
  \BibitemOpen
  \bibfield  {author} {\bibinfo {author} {\bibfnamefont {N.}~\bibnamefont {Abgrall}} \emph {et~al.} (\bibinfo {collaboration} {LEGEND collaboration}),\ }\href {https://doi.org/10.1063/1.5007652} {\bibfield  {journal} {\bibinfo  {journal} {AIP Conf. Proc.}\ }\textbf {\bibinfo {volume} {1894}},\ \bibinfo {pages} {020027} (\bibinfo {year} {2017})},\ \Eprint {https://arxiv.org/abs/1709.01980} {arXiv:1709.01980 [physics.ins-det]} \BibitemShut {NoStop}%
\bibitem [{\citenamefont {Adhikari}\ \emph {et~al.}(2021)\citenamefont {Adhikari} \emph {et~al.}}]{nexo}%
  \BibitemOpen
  \bibfield  {author} {\bibinfo {author} {\bibfnamefont {G.}~\bibnamefont {Adhikari}} \emph {et~al.} (\bibinfo {collaboration} {nEXO Collaboration}),\ }\href {https://doi.org/10.1088/1361-6471/ac3631} {\bibfield  {journal} {\bibinfo  {journal} {Journal of Physics G: Nuclear and Particle Physics}\ }\textbf {\bibinfo {volume} {49}},\ \bibinfo {pages} {015104} (\bibinfo {year} {2021})}\BibitemShut {NoStop}%
\bibitem [{\citenamefont {Alfonso}\ \emph {et~al.}(2022)\citenamefont {Alfonso} \emph {et~al.}}]{cupid}%
  \BibitemOpen
  \bibfield  {author} {\bibinfo {author} {\bibfnamefont {K.}~\bibnamefont {Alfonso}} \emph {et~al.} (\bibinfo {collaboration} {CUPID Collaboration}),\ }\bibfield  {journal} {\bibinfo  {journal} {Journal of Low Temperature Physics}\ }\textbf {\bibinfo {volume} {211}},\ \href {https://doi.org/10.1007/s10909-022-02909-3} {10.1007/s10909-022-02909-3} (\bibinfo {year} {2022})\BibitemShut {NoStop}%
\bibitem [{\citenamefont {Avignone}\ \emph {et~al.}(2005)\citenamefont {Avignone}, \citenamefont {King},\ and\ \citenamefont {Zdesenko}}]{sensitivity_bi}%
  \BibitemOpen
  \bibfield  {author} {\bibinfo {author} {\bibfnamefont {F.~T.}\ \bibnamefont {Avignone}}, \bibinfo {author} {\bibfnamefont {G.~S.}\ \bibnamefont {King}},\ and\ \bibinfo {author} {\bibfnamefont {Y.~G.}\ \bibnamefont {Zdesenko}},\ }\href {https://doi.org/10.1088/1367-2630/7/1/006} {\bibfield  {journal} {\bibinfo  {journal} {New Journal of Physics}\ }\textbf {\bibinfo {volume} {7}},\ \bibinfo {pages} {6} (\bibinfo {year} {2005})}\BibitemShut {NoStop}%
\bibitem [{\citenamefont {Avignone}\ \emph {et~al.}(2008)\citenamefont {Avignone}, \citenamefont {Elliott},\ and\ \citenamefont {Engel}}]{review}%
  \BibitemOpen
  \bibfield  {author} {\bibinfo {author} {\bibfnamefont {F.~T.}\ \bibnamefont {Avignone}}, \bibinfo {author} {\bibfnamefont {S.~R.}\ \bibnamefont {Elliott}},\ and\ \bibinfo {author} {\bibfnamefont {J.}~\bibnamefont {Engel}},\ }\href {https://doi.org/10.1103/RevModPhys.80.481} {\bibfield  {journal} {\bibinfo  {journal} {Rev. Mod. Phys.}\ }\textbf {\bibinfo {volume} {80}},\ \bibinfo {pages} {481} (\bibinfo {year} {2008})}\BibitemShut {NoStop}%
\bibitem [{\citenamefont {Abgrall}\ \emph {et~al.}(2018)\citenamefont {Abgrall} \emph {et~al.}}]{mjd_enr}%
  \BibitemOpen
  \bibfield  {author} {\bibinfo {author} {\bibfnamefont {N.}~\bibnamefont {Abgrall}} \emph {et~al.} (\bibinfo {collaboration} {MAJORANA}),\ }\href {https://doi.org/10.1016/j.nima.2017.09.036} {\bibfield  {journal} {\bibinfo  {journal} {Nucl. Instrum. Meth. A}\ }\textbf {\bibinfo {volume} {877}},\ \bibinfo {pages} {314} (\bibinfo {year} {2018})},\ \Eprint {https://arxiv.org/abs/1707.06255} {arXiv:1707.06255 [physics.ins-det]} \BibitemShut {NoStop}%
\bibitem [{\citenamefont {Avignone~III}\ and\ \citenamefont {Elliott}(2019)}]{bb_ge}%
  \BibitemOpen
  \bibfield  {author} {\bibinfo {author} {\bibfnamefont {F.~T.}\ \bibnamefont {Avignone~III}}\ and\ \bibinfo {author} {\bibfnamefont {S.~R.}\ \bibnamefont {Elliott}},\ }\href {https://doi.org/10.3389/fphy.2019.00006} {\bibfield  {journal} {\bibinfo  {journal} {Frontiers in Physics}\ }\textbf {\bibinfo {volume} {7}},\ \bibinfo {pages} {6} (\bibinfo {year} {2019})}\BibitemShut {NoStop}%
\bibitem [{\citenamefont {Rahaman}\ \emph {et~al.}(2008)\citenamefont {Rahaman}, \citenamefont {Elomaa}, \citenamefont {Eronen}, \citenamefont {Hakala}, \citenamefont {Jokinen}, \citenamefont {Julin}, \citenamefont {Kankainen}, \citenamefont {Saastamoinen}, \citenamefont {Suhonen}, \citenamefont {Weber},\ and\ \citenamefont {Äystö}}]{q_val_1}%
  \BibitemOpen
  \bibfield  {author} {\bibinfo {author} {\bibfnamefont {S.}~\bibnamefont {Rahaman}}, \bibinfo {author} {\bibfnamefont {V.-V.}\ \bibnamefont {Elomaa}}, \bibinfo {author} {\bibfnamefont {T.}~\bibnamefont {Eronen}}, \bibinfo {author} {\bibfnamefont {J.}~\bibnamefont {Hakala}}, \bibinfo {author} {\bibfnamefont {A.}~\bibnamefont {Jokinen}}, \bibinfo {author} {\bibfnamefont {J.}~\bibnamefont {Julin}}, \bibinfo {author} {\bibfnamefont {A.}~\bibnamefont {Kankainen}}, \bibinfo {author} {\bibfnamefont {A.}~\bibnamefont {Saastamoinen}}, \bibinfo {author} {\bibfnamefont {J.}~\bibnamefont {Suhonen}}, \bibinfo {author} {\bibfnamefont {C.}~\bibnamefont {Weber}},\ and\ \bibinfo {author} {\bibfnamefont {J.}~\bibnamefont {Äystö}},\ }\href {https://doi.org/https://doi.org/10.1016/j.physletb.2008.02.047} {\bibfield  {journal} {\bibinfo  {journal} {Physics Letters B}\ }\textbf {\bibinfo {volume} {662}},\ \bibinfo {pages} {111} (\bibinfo {year} {2008})}\BibitemShut {NoStop}%
\bibitem [{\citenamefont {Douysset}\ \emph {et~al.}(2001)\citenamefont {Douysset}, \citenamefont {Fritioff}, \citenamefont {Carlberg}, \citenamefont {Bergstr{\"o}m},\ and\ \citenamefont {Bj{\"o}rkhage}}]{q_val_2}%
  \BibitemOpen
  \bibfield  {author} {\bibinfo {author} {\bibfnamefont {G.}~\bibnamefont {Douysset}}, \bibinfo {author} {\bibfnamefont {T.}~\bibnamefont {Fritioff}}, \bibinfo {author} {\bibfnamefont {C.}~\bibnamefont {Carlberg}}, \bibinfo {author} {\bibfnamefont {I.}~\bibnamefont {Bergstr{\"o}m}},\ and\ \bibinfo {author} {\bibfnamefont {M.}~\bibnamefont {Bj{\"o}rkhage}},\ }\href {https://doi.org/10.1103/PhysRevLett.86.4259} {\bibfield  {journal} {\bibinfo  {journal} {Phys. Rev. Lett.}\ }\textbf {\bibinfo {volume} {86}},\ \bibinfo {pages} {4259} (\bibinfo {year} {2001})}\BibitemShut {NoStop}%
\bibitem [{\citenamefont {Abgrall}\ \emph {et~al.}(2016{\natexlab{a}})\citenamefont {Abgrall} \emph {et~al.}}]{mjd_assay}%
  \BibitemOpen
  \bibfield  {author} {\bibinfo {author} {\bibfnamefont {N.}~\bibnamefont {Abgrall}} \emph {et~al.} (\bibinfo {collaboration} {\textsc{Majorana} Collaboration}),\ }\href {https://doi.org/10.1016/j.nima.2016.04.070} {\bibfield  {journal} {\bibinfo  {journal} {Nuclear Instruments and Methods in Physics Research Section A: Accelerators, Spectrometers, Detectors and Associated Equipment}\ }\textbf {\bibinfo {volume} {828}},\ \bibinfo {pages} {22} (\bibinfo {year} {2016}{\natexlab{a}})}\BibitemShut {NoStop}%
\bibitem [{\citenamefont {Alduino}\ \emph {et~al.}(2017)\citenamefont {Alduino} \emph {et~al.}}]{cuore_bgm}%
  \BibitemOpen
  \bibfield  {author} {\bibinfo {author} {\bibfnamefont {C.}~\bibnamefont {Alduino}} \emph {et~al.} (\bibinfo {collaboration} {Cuore Collaboration}),\ }\href {https://doi.org/10.1140/epjc/s10052-017-5080-6} {\bibfield  {journal} {\bibinfo  {journal} {The European Physical Journal C}\ }\textbf {\bibinfo {volume} {77}},\ \bibinfo {pages} {543} (\bibinfo {year} {2017})}\BibitemShut {NoStop}%
\bibitem [{\citenamefont {Albert}\ \emph {et~al.}(2018)\citenamefont {Albert} \emph {et~al.}}]{nexo_bgm}%
  \BibitemOpen
  \bibfield  {author} {\bibinfo {author} {\bibfnamefont {J.~B.}\ \bibnamefont {Albert}} \emph {et~al.} (\bibinfo {collaboration} {nEXO Collaboration}),\ }\href {https://doi.org/10.1103/PhysRevC.97.065503} {\bibfield  {journal} {\bibinfo  {journal} {Phys. Rev. C}\ }\textbf {\bibinfo {volume} {97}},\ \bibinfo {pages} {065503} (\bibinfo {year} {2018})}\BibitemShut {NoStop}%
\bibitem [{\citenamefont {Tsang}\ \emph {et~al.}(2018)\citenamefont {Tsang}, \citenamefont {Arnquist}, \citenamefont {Hoppe}, \citenamefont {Orrell},\ and\ \citenamefont {Saldanha}}]{assaydistributions}%
  \BibitemOpen
  \bibfield  {author} {\bibinfo {author} {\bibfnamefont {R.~H.~M.}\ \bibnamefont {Tsang}}, \bibinfo {author} {\bibfnamefont {I.~J.}\ \bibnamefont {Arnquist}}, \bibinfo {author} {\bibfnamefont {E.~W.}\ \bibnamefont {Hoppe}}, \bibinfo {author} {\bibfnamefont {J.~L.}\ \bibnamefont {Orrell}},\ and\ \bibinfo {author} {\bibfnamefont {R.}~\bibnamefont {Saldanha}},\ }\href@noop {} {\  (\bibinfo {year} {2018})},\ \Eprint {https://arxiv.org/abs/1808.05307} {arXiv:1808.05307 [physics.data-an]} \BibitemShut {NoStop}%
\bibitem [{\citenamefont {Abgrall}\ \emph {et~al.}(2014)\citenamefont {Abgrall} \emph {et~al.}}]{mjd_design}%
  \BibitemOpen
  \bibfield  {author} {\bibinfo {author} {\bibfnamefont {N.}~\bibnamefont {Abgrall}} \emph {et~al.} (\bibinfo {collaboration} {\textsc{Majorana} Collaboration}),\ }\href {https://doi.org/10.1155/2014/365432} {\bibfield  {journal} {\bibinfo  {journal} {Advances in High Energy Physics}\ }\textbf {\bibinfo {volume} {2014}},\ \bibinfo {pages} {365432} (\bibinfo {year} {2014})}\BibitemShut {NoStop}%
\bibitem [{\citenamefont {Heise}(2015)}]{surf}%
  \BibitemOpen
  \bibfield  {author} {\bibinfo {author} {\bibfnamefont {J.}~\bibnamefont {Heise}},\ }\href {https://doi.org/10.1088/1742-6596/606/1/012015} {\bibfield  {journal} {\bibinfo  {journal} {Journal of Physics: Conference Series}\ }\textbf {\bibinfo {volume} {606}},\ \bibinfo {pages} {012015} (\bibinfo {year} {2015})}\BibitemShut {NoStop}%
\bibitem [{\citenamefont {Hoppe}\ \emph {et~al.}(2014)\citenamefont {Hoppe}, \citenamefont {Aalseth}, \citenamefont {Farmer}, \citenamefont {Hossbach}, \citenamefont {Liezers}, \citenamefont {Miley}, \citenamefont {Overman},\ and\ \citenamefont {Reeves}}]{ugefcu}%
  \BibitemOpen
  \bibfield  {author} {\bibinfo {author} {\bibfnamefont {E.}~\bibnamefont {Hoppe}}, \bibinfo {author} {\bibfnamefont {C.}~\bibnamefont {Aalseth}}, \bibinfo {author} {\bibfnamefont {O.}~\bibnamefont {Farmer}}, \bibinfo {author} {\bibfnamefont {T.}~\bibnamefont {Hossbach}}, \bibinfo {author} {\bibfnamefont {M.}~\bibnamefont {Liezers}}, \bibinfo {author} {\bibfnamefont {H.}~\bibnamefont {Miley}}, \bibinfo {author} {\bibfnamefont {N.}~\bibnamefont {Overman}},\ and\ \bibinfo {author} {\bibfnamefont {J.}~\bibnamefont {Reeves}},\ }\href {https://doi.org/https://doi.org/10.1016/j.nima.2014.06.082} {\bibfield  {journal} {\bibinfo  {journal} {Nuclear Instruments and Methods in Physics Research Section A: Accelerators, Spectrometers, Detectors and Associated Equipment}\ }\textbf {\bibinfo {volume} {764}},\ \bibinfo {pages} {116} (\bibinfo {year} {2014})}\BibitemShut {NoStop}%
\bibitem [{\citenamefont {Bugg}\ \emph {et~al.}(2014)\citenamefont {Bugg}, \citenamefont {Efremenko},\ and\ \citenamefont {Vasilyev}}]{muonveto}%
  \BibitemOpen
  \bibfield  {author} {\bibinfo {author} {\bibfnamefont {W.}~\bibnamefont {Bugg}}, \bibinfo {author} {\bibfnamefont {Y.}~\bibnamefont {Efremenko}},\ and\ \bibinfo {author} {\bibfnamefont {S.}~\bibnamefont {Vasilyev}},\ }\href {https://doi.org/https://doi.org/10.1016/j.nima.2014.05.055} {\bibfield  {journal} {\bibinfo  {journal} {Nuclear Instruments and Methods in Physics Research Section A: Accelerators, Spectrometers, Detectors and Associated Equipment}\ }\textbf {\bibinfo {volume} {758}},\ \bibinfo {pages} {91} (\bibinfo {year} {2014})}\BibitemShut {NoStop}%
\bibitem [{\citenamefont {Abgrall}\ \emph {et~al.}(2022)\citenamefont {Abgrall} \emph {et~al.}}]{lmfe}%
  \BibitemOpen
  \bibfield  {author} {\bibinfo {author} {\bibfnamefont {N.}~\bibnamefont {Abgrall}} \emph {et~al.} (\bibinfo {collaboration} {\textsc{Majorana} Collaboration}),\ }\href {https://doi.org/10.1088/1748-0221/17/05/T05003} {\bibfield  {journal} {\bibinfo  {journal} {Journal of Instrumentation}\ }\textbf {\bibinfo {volume} {17}}\bibinfo  {number} { (05)},\ \bibinfo {pages} {T05003}}\BibitemShut {NoStop}%
\bibitem [{\citenamefont {Abgrall}\ \emph {et~al.}(2016{\natexlab{b}})\citenamefont {Abgrall} \emph {et~al.}}]{cables}%
  \BibitemOpen
\bibfield  {number} {  }\bibfield  {author} {\bibinfo {author} {\bibfnamefont {N.}~\bibnamefont {Abgrall}} \emph {et~al.} (\bibinfo {collaboration} {\textsc{Majorana} Collaboration}),\ }\href {https://doi.org/https://doi.org/10.1016/j.nima.2016.04.006} {\bibfield  {journal} {\bibinfo  {journal} {Nuclear Instruments and Methods in Physics Research Section A: Accelerators, Spectrometers, Detectors and Associated Equipment}\ }\textbf {\bibinfo {volume} {823}},\ \bibinfo {pages} {83} (\bibinfo {year} {2016}{\natexlab{b}})}\BibitemShut {NoStop}%
\bibitem [{\citenamefont {Boswell}\ \emph {et~al.}(2011)\citenamefont {Boswell} \emph {et~al.}}]{mage}%
  \BibitemOpen
  \bibfield  {author} {\bibinfo {author} {\bibfnamefont {M.}~\bibnamefont {Boswell}} \emph {et~al.},\ }\href {https://doi.org/10.1109/TNS.2011.2144619} {\bibfield  {journal} {\bibinfo  {journal} {IEEE Transactions on Nuclear Science}\ }\textbf {\bibinfo {volume} {58}},\ \bibinfo {pages} {1212} (\bibinfo {year} {2011})}\BibitemShut {NoStop}%
\bibitem [{\citenamefont {Alvis}\ \emph {et~al.}(2019)\citenamefont {Alvis} \emph {et~al.}}]{mjd_26}%
  \BibitemOpen
  \bibfield  {author} {\bibinfo {author} {\bibfnamefont {S.~I.}\ \bibnamefont {Alvis}} \emph {et~al.} (\bibinfo {collaboration} {\textsc{Majorana} Collaboration}),\ }\href {https://doi.org/10.1103/PhysRevC.100.025501} {\bibfield  {journal} {\bibinfo  {journal} {Phys. Rev. C}\ }\textbf {\bibinfo {volume} {100}},\ \bibinfo {pages} {025501} (\bibinfo {year} {2019})}\BibitemShut {NoStop}%
\bibitem [{\citenamefont {Gilliss}(2019)}]{gilliss_thesis}%
  \BibitemOpen
  \bibfield  {author} {\bibinfo {author} {\bibfnamefont {T.~F.}\ \bibnamefont {Gilliss}},\ }\href {http://libproxy.lib.unc.edu/login?url=https://www.proquest.com/dissertations-theses/statistical-modeling-markov-chain-monte-carlo/docview/2311051940/se-2?accountid=14244} {\bibfield  {journal} {\bibinfo  {journal} {ProQuest Dissertations and Theses}\ ,\ \bibinfo {pages} {165}} (\bibinfo {year} {2019})}\BibitemShut {NoStop}%
\bibitem [{\citenamefont {Agostini}\ \emph {et~al.}(2017)\citenamefont {Agostini} \emph {et~al.}}]{gerda_germanium_assay}%
  \BibitemOpen
  \bibfield  {author} {\bibinfo {author} {\bibfnamefont {M.}~\bibnamefont {Agostini}} \emph {et~al.},\ }\bibfield  {title} {\bibinfo {title} {Limits on uranium and thorium bulk content in gerda phase i detectors},\ }\href {https://doi.org/https://doi.org/10.1016/j.astropartphys.2017.03.003} {\bibfield  {journal} {\bibinfo  {journal} {Astroparticle Physics}\ }\textbf {\bibinfo {volume} {91}},\ \bibinfo {pages} {15} (\bibinfo {year} {2017})}\BibitemShut {NoStop}%
\bibitem [{\citenamefont {Workman}\ \emph {et~al.}(2022)\citenamefont {Workman} \emph {et~al.}}]{PDG}%
  \BibitemOpen
  \bibfield  {author} {\bibinfo {author} {\bibfnamefont {R.~L.}\ \bibnamefont {Workman}} \emph {et~al.} (\bibinfo {collaboration} {Particle Data Group}),\ }\href {https://doi.org/10.1093/ptep/ptac097} {\bibfield  {journal} {\bibinfo  {journal} {PTEP}\ }\textbf {\bibinfo {volume} {2022}},\ \bibinfo {pages} {083C01} (\bibinfo {year} {2022})}\BibitemShut {NoStop}%
\end{thebibliography}
\end{document}